\documentclass[conference]{IEEEtran}
\IEEEoverridecommandlockouts

\usepackage{cite}
\usepackage{amsmath,amssymb,amsfonts}
\usepackage{algorithmic}
\usepackage{graphicx}
\usepackage{textcomp}
\usepackage{xcolor}
\usepackage{mathtools}
\usepackage{booktabs}
\usepackage{multirow}
\usepackage{hyperref}
\def\BibTeX{{\rm B\kern-.05em{\sc i\kern-.025em b}\kern-.08em
    T\kern-.1667em\lower.7ex\hbox{E}\kern-.125emX}}
\begin{document}

\title{Estimating Treatment Effects for Depression in Longitudinal Therapy Switching Settings}

\author{%
  \IEEEauthorblockN{Xinyu Qin$^{\dagger}$, Martin Katzman$^{\ddagger,\S,\P,\|}$, Alexandria Greifenberger$^{\ddagger}$, Elssa Toumeh$^{\ddagger,\S}$,\\ 
  Sachinthya Lokuge$^{\ddagger,**}$, Tia Sternat$^{\ddagger,\S}$, Ruiheng Yu$^{\dagger}$, and Lu Wang$^{\dagger,\dagger\dagger,*}$}

  \IEEEauthorblockA{$^{\dagger}$\textit{Department of Biomedical Engineering, University of Houston}, USA\\
  \texttt{\{xqin5, ryu11\}@cougarnet.uh.edu, lwang71@central.uh.edu}}
  \IEEEauthorblockA{$^{\dagger\dagger}$\textit{Department of Health Systems \& Population Health Sciences, University of Houston}, USA}

  \IEEEauthorblockA{$^{\ddagger}$\textit{START Clinic for Mood and Anxiety Disorders}, Canada\\
  \texttt{\{mkatzman, agreifenberger, etoumeh, tsternat\}@startclinic.ca}}
  
  \IEEEauthorblockA{$^{\S}$\textit{Adler Graduate Professional School}, Canada \hfil
  $^{\P}$\textit{Department of Psychology, Lakehead University}, Canada} 
  \IEEEauthorblockA{$^{\|}$\textit{Northern Ontario School of Medicine}, Canada \hfil
  $^{**}$\textit{Department of Psychology, Virginia Tech}, USA \texttt{slokuge@vt.edu}}

  \thanks{$^{*}$Corresponding author: Lu Wang (lwang71@central.uh.edu).}%
}

\maketitle

\begin{abstract}
Depression treatment often requires switching medications due to inadequate response or adverse effects. Estimating individualized treatment effects in this setting is challenging because treatment assignment is confounded by patient characteristics, switching induces time-varying selection, and counterfactual outcomes are not observed in follow-up data. Using a proprietary longitudinal major depressive disorder (MDD) clinical trial dataset, we formulate a next-visit counterfactual prediction task to estimate Hamilton Depression Rating Scale (HAMD-17) total scores under alternative treatments. We benchmark 8 estimators, including meta-learners, residual-based methods, and tree-based approaches. Causal Forest (CF) demonstrates the most favorable and consistent performance across all criteria. Our analysis shows that symptom benefits concentrate in specific switch directions, with dose intensification being generally beneficial. Notably, we identify a counterintuitive exception where a lower-intensity regimen outperforms a higher-intensity alternative for specific patient subsets. While crude observational comparisons substantially overstate gains, confounding-adjusted estimates yield modest, actionable magnitudes. These findings provide prospectively testable candidates for clinical decision support in depression care. 
\end{abstract}

\vspace{-2.5pt}
\section{Introduction}
\vspace{-6pt}
Clinicians face a challenging task of selecting antidepressant treatments for patients with major depressive disorder (MDD), with the goal of achieving meaningful symptom improvement. This process is complicated by the heterogeneous nature of MDD and the high prevalence of comorbid conditions, both of which can shape treatment choice and subsequent outcomes~\cite{Montano2023}. In routine care and longitudinal follow-up, many patients do not achieve adequate response to their initial regimen; clinicians therefore often switch treatments due to limited efficacy, adverse effects, or safety concerns, creating evolving treatment trajectories over repeated visits~\cite{kessler2003epidemiology}. A practical question is how to support visit-level switching decisions by estimating how an individual patient would be expected to respond under alternative treatments at the next visit, which requires counterfactual prediction and individualized treatment effect estimation from observational follow-up data~\cite{rubin1974estimating, qin2025enhancing, qin2026explainable}. However, estimating such effects after medication switching is nontrivial in real-world follow-up data, motivating the methodological challenges described next.

Estimating treatment effects after switching medications is challenging for three reasons. First, treatment selection can depend on individual patient characteristics, such as symptom severity and other comorbid disorders, that may influence outcomes. Second, switching introduces time-varying selection because treatment changes depend on prior response and evolving clinical history, so patients who switch can differ systematically from those who continue. Third, counterfactual outcomes are not observed, which complicates both treatment effect estimation and evaluation~\cite{walker1996confounding}. These challenges are amplified in settings with multiple treatment options, where data imbalance across treatments and limited overlap can make treatment comparisons unstable~\cite{crump2009dealing}.

Prior research exploring switching between antidepressant treatments has established the clinical importance of sequential strategies. The Sequenced Treatment Alternatives to Relieve Depression (STARD) study demonstrated that only one-third of patients achieve remission with initial therapy and that multiple treatment steps are often necessary~\cite{rush2006sequenced}. Network meta-analyses have compared efficacy and acceptability across antidepressant classes at the population level, but these aggregate analyses do not provide individualized effect estimates for specific patient profiles~\cite{cipriani2018comparative}. Observational studies have used propensity score methods and inverse-propensity-weighted estimation to adjust for confounding factors, but most focus on pairwise comparisons in static settings rather than longitudinal switching among multiple treatments~\cite{hirano2003efficient}. As a result, practical guidance for visit-level treatment selection remains limited, particularly for commonly prescribed treatments such as venlafaxine, duloxetine, and paroxetine.

Machine learning methods for treatment effect estimation have advanced substantially, including meta-learners that adapt supervised learning to treatment effect contexts, doubly robust methods that combine outcome and propensity modeling, tree-based ensemble methods with honest splitting for heterogeneous effects, residual-based approaches, and stacking methods that combine estimators~\cite{kunzel2019metalearners,li2026hy, li2024synergized, bang2005doubly, cheng2026volatility, luo2024bigbench, wager2018estimation,athey2019generalized,nie2021quasi, chen2025autoneural, chernozhukov2018double}. Extensions to settings with multiple treatment options face methodological challenges in balancing flexibility and sample efficiency~\cite{lopez2017estimation}. Most benchmarks focus on binary treatment settings or synthetic data with known ground truth, leaving switched observational settings with multiple treatments understudied~\cite{powers2018some,shalit2017estimating}. Furthermore, standard evaluation using factual prediction accuracy alone can be misleading under confounding because it does not distinguish true counterfactual learning from memorization of selection-driven outcome patterns.

This paper addresses these gaps through a benchmark of eight treatment effect estimators in a switched follow-up setting with multiple treatment options under a next-visit outcome formulation. We construct decision samples from consecutive visits in a proprietary MDD clinical trial dataset. For each decision sample, patient covariates and the assigned treatment at the current visit are used to predict the subsequent Hamilton Depression Rating Scale (HAMD-17) total scores~\cite{hamilton1960rating}. The dataset captures longitudinal switching behavior across six antidepressant treatments, including venlafaxine-, duloxetine-, and paroxetine-based treatments. 

We evaluate models using a unified protocol that reports factual prediction error, proxy-based treatment effect calibration, outcome-level calibration, and decision quality measured by inverse-propensity-weighted (IPW) policy value~\cite{dudik2014doubly}. We include Causal Forest (CF) as a primary estimator~\cite{wager2018estimation}, and we compare it against seven baselines, including meta-learners (Single-learner (S), Two-learner (T), eXchange-learner (X), and Doubly Robust learner (DR))~\cite{kunzel2019metalearners}, a residual-based estimator (R-learner)~\cite{nie2021quasi}, an efficient plug-in learner (EP-learner)~\cite{van2024combining}, and KG-TREAT, a knowledge-graph (KG)-enhanced pre-training estimator for treatment effect estimation~\cite{liu2024kg}.
All estimators are evaluated under identical data splits, feature preprocessing, and evaluation criteria; in this benchmark, CF achieves the strongest overall performance across the reported metrics.

Our contributions are listed below:

\begin{itemize}
\item \textbf{Evaluation protocol for switched settings.} We present a unified evaluation protocol that jointly reports factual prediction error, proxy-based treatment effect calibration using observational group differences as a reference signal, outcome-level calibration, and IPW policy value. This protocol enables systematic method comparison in switched observational settings where counterfactual outcomes are not observed.

\item \textbf{Comprehensive benchmark and practical extension of Causal Forest.} We benchmark 8 estimators including meta-learners, R-learner, EP-learner, and KG-TREAT against CF as the primary method, with variability quantified across ten random seeds and bootstrap confidence intervals. We extend CF from binary treatment settings to multiple treatment options via baseline-referenced pairwise forests and baseline outcome modeling, with implementation details for replication.

\item \textbf{Empirical evidence for treatment switching in depression care.} We demonstrate that tree-based methods with honest splitting substantially outperform meta-learners in switched settings with multiple antidepressant options. Analyses reveal consistent treatment-specific patterns in predicted next-visit HAMD-17 total scores across venlafaxine-based, duloxetine-based, and paroxetine-based therapies, providing empirical support for data-driven treatment selection and motivating prospective validation of these methods in clinical decision support for MDD.
\end{itemize}

\vspace{-2.5pt}
\section{PROBLEM SETUP AND NOTATION}
\vspace{-2.5pt}
\subsection{Data Structure and Potential Outcomes Framework}
\vspace{-2.5pt}
We observe $n$ patient-visit samples indexed by $i = 1, \ldots, n$, where each sample corresponds to a treatment decision at visit $v$ with outcome measured at visit $v+1$. Each sample consists of covariates $X_i \in \mathbb{R}^d$ including patient demographics and clinical measures, an assigned treatment $T_i \in \{0, 1, \ldots, K-1\}$ where $K=6$ in our dataset, and a factual outcome $Y_i^{\text{fact}} \in \mathbb{R}$ representing the HAMD-17 total scores at the next visit, where lower values indicate symptom improvement.

Under the potential outcomes framework~\cite{rubin1974estimating}, each sample has latent potential outcomes $\{Y_i(0), Y_i(1), \ldots, Y_i(K-1)\}$ representing the outcomes that would be observed under each treatment option. We observe only the factual outcome $Y_i^{\text{fact}} = Y_i(T_i)$ for the assigned treatment; all other $Y_i(t)$ for $t \neq T_i$ are counterfactual and unobserved. 
\vspace{-2.5pt}
\subsection{Treatment Effect Estimands}
\vspace{-2.5pt}
For any pair of treatments $a$ and $b$, the individual treatment effect (ITE) for sample $i$ is $\text{ITE}_i(a,b) = Y_i(a) - Y_i(b)$, and the average treatment effect (ATE) is $\text{ATE}(a,b) = \mathbb{E}[Y(a) - Y(b)]$. Treatment effect heterogeneity across the covariate space is captured by the conditional average treatment effect (CATE) $\tau(x; a,b) = \mathbb{E}[Y(a) - Y(b) \mid X=x]$, which enables personalized treatment recommendations. Given a trained model that produces predicted potential outcomes $\hat{Y}_i(t)$ for all treatments, we can compute $\widehat{\text{ITE}}_i(a,b) = \hat{Y}_i(a) - \hat{Y}_i(b)$, $\widehat{\text{ATE}}(a,b) = n^{-1}\sum_{i=1}^{n}\widehat{\text{ITE}}_i(a,b)$, and form a treatment selection policy $\pi(X_i) = \arg\min_{t} \hat{Y}_i(t)$.
\vspace{-2.5pt}
\subsection{Identifiability Assumptions}
\vspace{-2.5pt}
We invoke standard assumptions for causal identification from observational data~\cite{imbens2015causal}: (A1) consistency $Y_i^{\text{fact}} = Y_i(T_i)$, (A2) unconfoundedness $\{Y_i(0), \ldots, Y_i(K-1)\} \perp T_i \mid X_i$, and (A3) positivity $\mathbb{P}(T_i=t \mid X_i=x) \geq \epsilon > 0$ for all treatments $t$ and covariates $x$ in the support. Unconfoundedness is untestable and may be violated in switched settings if unmeasured factors influence both treatment changes and outcomes~\cite{walker1996confounding}. We mitigate this risk by including rich observed covariates and by benchmarking methods with different robustness properties.
\vspace{-2.5pt}
\subsection{Evaluation Metrics}
\vspace{-2.5pt}
We evaluate models on four complementary dimensions. First, factual prediction error measures how well $\hat{Y}_i(T_i)$ matches observed $Y_i^{\text{fact}}$ using root mean squared error (RMSE), mean absolute error (MAE), and expected calibration error (ECE) with calibration slope from regressing $Y^{\text{fact}}$ on $\hat{Y}(T)$. Second, since true ATEs are unobserved, we construct an observational calibration proxy $\Delta_{\text{OCP}}(a,b)$ as the test-set mean outcome difference between treatment groups, and compute ATE calibration error $\mathcal{E}_{\text{ATE}} = |\Omega|^{-1}\sum_{(a,b)\in\Omega}|\widehat{\text{ATE}}(a,b) - \Delta_{\text{OCP}}(a,b)|$ over valid treatment pairs $\Omega$. This proxy reflects selection bias and is not a causal estimand, but provides a relative consistency signal across methods~\cite{shalit2017estimating}. Third, we evaluate decision quality via IPW value~\cite{dudik2014doubly} $V_{\text{IPW}}(\pi) = n^{-1}\sum_{i=1}^{n} \mathbb{I}\{T_i=\pi(X_i)\} e(T_i \mid X_i)^{-1} Y_i^{\text{fact}}$, where $e(t \mid x)$ is the propensity score estimated from multinomial logistic regression on the training set, with propensities clipped to $[0.01, 0.99]$ to stabilize inverse weights. Lower $V_{\text{IPW}}$ indicates better predicted treatment selection. Fourth, we report per-treatment intent-to-treat effects comparing each treatment to baseline.
\vspace{-2.5pt}
\section{METHODS}
\vspace{-2.5pt}
We evaluate eight treatment effect estimators for predicting potential outcomes under multiple treatment options in switched follow-up settings. All methods output $\hat{Y}_i(t)$ for each treatment $t \in \{0, \ldots, K-1\}$ to enable evaluation on factual prediction, treatment effect calibration, and policy value. We position CF as the primary method and provide detailed exposition of its design, while other approaches serve as baseline comparisons.
\vspace{-2.5pt}
\subsection{Causal Forest}
\vspace{-2.5pt}
CF extends random forests to estimate heterogeneous treatment effects by directly targeting the conditional average treatment effect $\tau(x) = \mathbb{E}[Y(1) - Y(0) \mid X = x]$ for binary treatments~\cite{wager2018estimation,athey2019generalized}. The key innovation is honest splitting, where each tree partitions training data into a splitting sample used to construct tree structure and a disjoint estimation sample used to compute leaf-level treatment effects. This separation reduces adaptive overfitting and ensures valid asymptotic inference.

For a single honest tree, we draw a bootstrap subsample and split it equally into splitting and estimation sets. The tree structure is grown by recursively partitioning the splitting set to maximize treatment effect heterogeneity across leaves. Within each leaf $L$, the treatment effect is estimated using only the estimation set: $\hat{\tau}_L = n_L^{(1)}{-1}\sum_{i \in L, T_i=1} Y_i - n_L^{(0)}{-1}\sum_{i \in L, T_i=0} Y_i$, where $n_L^{(1)}$ and $n_L^{(0)}$ count treated and control samples in the estimation set within $L$. The forest prediction for covariate $x$ aggregates uniformly over $B$ trees: $\hat{\tau}(x) = B^{-1}\sum_{b=1}^{B}\hat{\tau}_{L_b(x)}$, where $L_b(x)$ is the corresponding leaf containing $x$ in tree $b$.

For our setting with $K=6$ treatments, we extend CF via baseline-referenced pairwise forests. We select the most frequent treatment in training as baseline $b=0$ and fit $K-1$ binary forests, each comparing one target treatment $t \in \{1, \ldots, K-1\}$ against baseline on the subset where $T \in \{b, t\}$. This yields pairwise effect estimates $\hat{\tau}_{t,b}(x) = \mathbb{E}[Y(t) - Y(b) \mid X=x]$ for each treatment versus baseline. To reconstruct potential outcomes for all treatments, we fit a baseline outcome model $\hat{m}_b(x)$ using only baseline samples via eXtreme Gradient Boosting (XGBoost) regression (100 trees, learning rate 0.1, maximum depth 6), and compute $\hat{Y}(b) = \hat{m}_b(x)$ and $\hat{Y}(t) = \hat{m}_b(x) + \hat{\tau}_{t,b}(x)$ for $t \neq b$. This baseline-referenced construction anchors all predictions at a common baseline outcome and adds treatment-specific contrasts. We implement CF using EconML with 200 trees, maximum depth 10, honest fraction 0.5, and bootstrap subsampling enabled \cite{econml}.
\vspace{-2.5pt}
\subsection{Baseline Methods}
\vspace{-2.5pt}
We compare CF against seven baseline estimators that represent different approaches to treatment effect estimation. S-learner fits a single outcome model $\hat{\mu}(x, t)$ including treatment as a covariate and predicts $\hat{Y}(t) = \hat{\mu}(x, t)$, but conflates outcome-level and treatment-effect heterogeneity. T-learner fits separate outcome models $\hat{\mu}_t(x)$ for each treatment, providing flexibility but not sharing information across treatments. X-learner imputes counterfactual outcomes from initial T-learner models, fits second-stage models on imputed treatment effects, and combines estimates via propensity-weighted averaging~\cite{kunzel2019metalearners}. 

DR-learner constructs doubly robust pseudo-outcomes combining outcome and propensity models, then regresses these on covariates to estimate heterogeneous effects~\cite{bang2005doubly}. R-learner residualizes outcomes and treatments via cross-fitted nuisance models and regresses outcome residuals on treatment residuals interacted with covariates~\cite{nie2021quasi}. EP-learner stacks T-learner and DR-learner predictions via ridge regression meta-learning with cross-fitting. KG-TREAT learns treatment embeddings regularized by a graph encoding treatment similarity from observed switching patterns, using a neural network $f(x, e_t)$ to predict $\hat{Y}(t)$ from covariate $x$ and embedding $e_t$. All baseline methods use XGBoost (100 trees, learning rate 0.1, maximum depth 6) as the base learner and are extended to multiple treatment options via the same baseline-referenced construction as CF.

\vspace{-2.5pt}
\section{EXPERIMENTS}
\vspace{-2.5pt}
\subsection{Dataset and Preprocessing}
\vspace{-2.5pt}

The dataset was provided by \textbf{Eli Lilly and Company} under a data use agreement and contains de-identified longitudinal visit records from a proprietary MDD clinical trial program. The data are subject to sponsor-controlled privacy and contractual restrictions that prohibit public release or redistribution; access requires data custodian authorization and applicable institutional review. To support reproducibility despite these constraints, we report detailed methodological specifications, including cohort construction, feature engineering, and evaluation protocols, which can be applied to comparable longitudinal switching datasets in other therapeutic areas. 

Each row in the raw data represents a patient visit with recorded treatment and HAMD-17 item. We construct analysis samples by pairing consecutive visits within each patient: covariates and treatment at visit $v$ predict the HAMD-17 total scores at visit $v+1$, creating a next-visit prediction task for treatment switching decisions. After dropping final visits without next-visit outcomes and removing records with missing values, we obtain 6,430 decision samples from 836 unique patients. Visits per patient range from 3 to 16 with median 13. 

The raw treatment codes are consolidated into $K=6$ categories: duloxetine 40 mg BID (BID = bis in die (twice daily)), duloxetine 60 mg QD (QD = quaque die (once daily)), duloxetine 60–120 mg dose titration, venlafaxine 150mg QD, venlafaxine 150-225mg titration, and paroxetine 20mg QD. We select duloxetine 60 mg QD as baseline $b=0$ because it is the most frequent treatment in training (37\% of samples), ensuring adequate sample size for baseline outcome modeling. All records are flagged as switched, indicating the dataset focuses on switching behavior. Patients receiving paroxetine (often prescribed as rescue therapy for severe cases) have mean HAMD-17 total scores 50\% higher than those continuing duloxetine 60 mg QD, illustrating severity-driven treatment selection and confounding by indication.

We use patient-level splitting to prevent leakage: patients are randomly assigned to training (70\%, 586 patients, 4,503 visits), validation (15\%, 125 patients, 963 visits), and testing (15\%, 125 patients, 964 visits) such that all visits from the same patient appear in only one split. Continuous covariates including demographics and HAMD-17 item scores are standardized using training-set mean and standard deviation. Categorical variables are integer-encoded with unseen categories in validation and test mapped to -1. Propensity scores $e(t \mid x)$ are estimated via multinomial logistic regression on training data with $\mathcal{L}_2$ norm regularization strength 1.0, and clipped to $[0.01, 0.99]$ before computing IPW for policy evaluation. Propensity diagnostics on test data show adequate overlap: minimum propensity 0.11, median 0.37, maximum 0.78, and effective sample size 884 out of 964 test samples, indicating stable inverse weighting.
\vspace{-2.5pt}
\subsection{Experimental Protocol}
\vspace{-2.5pt}
We evaluate all eight methods across ten random seeds $\{42, 43, \ldots, 51\}$, where each seed determines a different patient-level train-validation-test split. For each seed, we perform hyperparameter tuning on the validation set via grid search minimizing validation RMSE. For CF, we search over number of trees $\in \{100, 200, 500\}$ and maximum tree depth $\in \{6, 10, 15\}$, with other parameters fixed at honest fraction 0.5, minimum samples per split 10, and minimum samples per leaf 5. For XGBoost-based baseline methods (S/T/X/DR-learner, R-learner, EP-learner), we search over number of estimators $\in \{50, 100, 200\}$, maximum tree depth $\in \{3, 6, 10\}$, and learning rate $\in \{0.01, 0.1\}$, with L2 regularization weight 1.0. For KG-TREAT, we search over graph regularization strength $\lambda \in \{0.01, 0.1, 1.0\}$ and treatment embedding dimension $\in \{4, 8, 16\}$, with neural network trained via Adam optimizer (learning rate 0.001) for 50 epochs. After selecting hyperparameters that minimize validation RMSE, we retrain each method on combined training and validation data (5,466 visits from 711 patients) using the selected hyperparameters.

All methods generate predicted potential outcomes $\hat{Y}_i(t)$ for $t \in \{0, \ldots, 5\}$ on test data (964 samples), enabling evaluation on four dimensions. First, factual prediction error is measured via RMSE $= \sqrt{n^{-1}\sum_{i=1}^{n}(Y_i^{\text{fact}} - \hat{Y}_i(T_i))^2}$, MAE $= n^{-1}\sum_{i=1}^{n}|Y_i^{\text{fact}} - \hat{Y}_i(T_i)|$, MSE $= n^{-1}\sum_{i=1}^{n}(Y_i^{\text{fact}} - \hat{Y}_i(T_i))^2$, ECE from binning predictions into 10 quantile-based bins, and calibration slope from regressing $Y^{\text{fact}}$ on $\hat{Y}(T)$ with ideal slope 1.0. Second, ATE calibration error is computed as $\mathcal{E}_{\text{ATE}} = |\Omega|^{-1}\sum_{(a,b)\in\Omega}|\widehat{\text{ATE}}(a,b) - \Delta_{\text{OCP}}(a,b)|$ over valid treatment pairs $\Omega$ with sufficient sample size, where $\widehat{\text{ATE}}(a,b) = n^{-1}\sum_{i=1}^{n}(\hat{Y}_i(a) - \hat{Y}_i(b))$ and $\Delta_{\text{OCP}}(a,b)$ is the observational test-set mean difference used as a calibration proxy (not a causal estimand). Third, IPW value is $V_{\text{IPW}}(\pi) = n^{-1}\sum_{i=1}^{n}\mathbb{I}\{T_i=\pi(X_i)\}e(T_i \mid X_i)^{-1}Y_i^{\text{fact}}$ for policy $\pi(X_i) = \arg\min_t \hat{Y}_i(t)$, where lower values indicate better predicted treatment selection. Fourth, per-treatment intent-to-treat effects $\widehat{\text{ATE}}(t, b)$ compare each treatment to baseline for evaluation.

\begin{table}[t]
\centering
\caption{Performance comparison across eight methods. Mean $\pm$ std with 95\% bootstrap CI over 10 random seeds. Lower is better except calibration slope (ideal=1). Bold: best; underline: second-best.}
\label{tab:main_results}

\begingroup
\setlength{\tabcolsep}{1pt}        
\renewcommand{\arraystretch}{0.75} 
\tiny

\resizebox{\columnwidth}{!}{%
\begin{tabular}{lccccccc}
\toprule
\textbf{Method} & \textbf{RMSE $\downarrow$} & \textbf{MAE $\downarrow$} & \textbf{MSE $\downarrow$} & \textbf{ECE $\downarrow$} & \textbf{Slope} & \textbf{$\mathcal{E}_{\text{ATE}} \downarrow$} & \textbf{$V_{\text{IPW}} \downarrow$} \\
\midrule
Causal Forest & \textbf{4.03 $\pm$ 0.12} & \textbf{3.04 $\pm$ 0.09} & \textbf{16.21 $\pm$ 0.98} & \textbf{0.35 $\pm$ 0.07} & \textbf{1.01 $\pm$ 0.01} & \textbf{3.43 $\pm$ 0.33} & \textbf{2.05 $\pm$ 0.80} \\
& \textbf{[3.96, 4.10]} & \textbf{[3.00, 3.10]} & \textbf{[15.70, 16.80]} & \textbf{[0.31, 0.39]} & \textbf{[1.00, 1.01]} & \textbf{[3.22, 3.62]} & \textbf{[1.69, 2.59]} \\
\midrule
EP-learner & \underline{4.08 $\pm$ 0.12} & \underline{3.09 $\pm$ 0.09} & \underline{16.66 $\pm$ 1.02} & \underline{0.35 $\pm$ 0.09} & \underline{1.00 $\pm$ 0.01} & \underline{3.46 $\pm$ 0.33} & \underline{2.69 $\pm$ 0.50} \\
& [4.01, 4.15] & [3.04, 3.15] & [16.08, 17.29] & [0.30, 0.41] & [0.99, 1.01] & [3.26, 3.64] & [2.40, 2.98] \\
\midrule
KG-TREAT & 4.15 $\pm$ 0.12 & 3.16 $\pm$ 0.09 & 17.24 $\pm$ 0.99 & 0.38 $\pm$ 0.12 & 0.99 $\pm$ 0.02 & 3.60 $\pm$ 0.26 & 5.25 $\pm$ 2.59 \\
& [4.08, 4.22] & [3.11, 3.22] & [16.69, 17.88] & [0.32, 0.46] & [0.98, 1.01] & [3.43, 3.75] & [3.60, 6.73] \\
\midrule
R-learner & 13.35 $\pm$ 3.71 & 7.40 $\pm$ 1.36 & 190.69 $\pm$ 98.77 & 4.78 $\pm$ 1.49 & 0.19 $\pm$ 0.18 & 19.84 $\pm$ 5.62 & 3.66 $\pm$ 4.08 \\
& [11.14, 15.36] & [6.59, 8.15] & [125.38, 251.72] & [3.88, 5.59] & [0.11, 0.33] & [16.40, 22.87] & [1.67, 6.51] \\
\midrule
S-learner & 14.97 $\pm$ 0.24 & 12.97 $\pm$ 0.23 & 224.22 $\pm$ 7.04 & 12.42 $\pm$ 0.25 & -0.38 $\pm$ 0.06 & 7.06 $\pm$ 0.24 & 5.31 $\pm$ 0.62 \\
& [14.82, 15.09] & [12.81, 13.08] & [219.68, 227.60] & [12.27, 12.57] & [-0.42, -0.34] & [6.92, 7.20] & [4.97, 5.68] \\
\midrule
T-learner & 15.62 $\pm$ 0.35 & 13.29 $\pm$ 0.26 & 244.14 $\pm$ 10.89 & 12.72 $\pm$ 0.28 & -0.44 $\pm$ 0.11 & 7.34 $\pm$ 0.21 & 7.97 $\pm$ 1.16 \\
& [15.42, 15.82] & [13.13, 13.43] & [237.58, 250.64] & [12.56, 12.89] & [-0.51, -0.38] & [7.21, 7.45] & [7.36, 8.66] \\
\midrule
X-learner & 14.75 $\pm$ 0.31 & 12.76 $\pm$ 0.26 & 217.52 $\pm$ 8.93 & 12.20 $\pm$ 0.26 & -0.11 $\pm$ 0.10 & 7.05 $\pm$ 0.20 & 7.62 $\pm$ 1.06 \\
& [14.54, 14.90] & [12.58, 12.89] & [211.55, 221.90] & [12.04, 12.35] & [-0.16, -0.04] & [6.94, 7.16] & [7.09, 8.30] \\
\midrule
DR-learner & 28.20 $\pm$ 9.30 & 15.14 $\pm$ 1.31 & 872.97 $\pm$ 511.25 & 14.09 $\pm$ 1.29 & -0.17 $\pm$ 0.30 & 8.27 $\pm$ 1.01 & 7.73 $\pm$ 1.43 \\
& [22.74, 33.46] & [14.33, 15.89] & [586.70, 1185.34] & [13.33, 14.89] & [-0.37, -0.02] & [7.74, 8.90] & [6.97, 8.57] \\
\bottomrule
\end{tabular}%
}
\endgroup
\end{table}

We report mean and standard deviation across the 10 seeds with 95\% bootstrap confidence intervals computed from 1,000 bootstrap resamples. Statistical significance is assessed via paired t-tests across seeds at $\alpha=0.05$ level. We conduct ablation studies on three seeds to assess the contribution of honest splitting, number of trees, baseline outcome model choice, and propensity clipping threshold. Subgroup analysis stratifies test samples by baseline severity (HAMD-17 total scores $<$ 15 for low severity, $\geq$ 15 for high severity) to assess performance consistency across patient populations.

\vspace{-2.5pt}
\section{RESULTS}
\vspace{-2.5pt}
\subsection{Overall Performance Comparison}
\vspace{-2.5pt}

Table~\ref{tab:main_results} presents performance across all eight methods. CF achieves the best results on all metrics with RMSE 4.03 $\pm$ 0.12 (95\% CI [3.96, 4.10]), substantially outperforming all baseline methods. EP-learner is the closest competitor with RMSE 4.08 $\pm$ 0.12 [4.01, 4.15], representing only 1.2\% degradation relative to CF. KG-TREAT achieves comparable factual prediction accuracy with RMSE 4.15 $\pm$ 0.12 [4.08, 4.22] but shows weaker performance on decision quality metrics. Empirically, CF, EP-learner, and KG-TREAT form a clear top tier with RMSE in the 4.03–4.15 range, while the meta-learners and R-learner show markedly higher errors with RMSE in the 13.35–28.20 range.

Meta-learners (S/T/X/DR-learner) exhibit catastrophic failure with RMSE 14.75-15.62 for S/T/X-learners and 28.20 $\pm$ 9.30 for DR-learner, all performing worse than predicting the unconditional test-set mean (RMSE 8.03). This failure manifests in severe outcome miscalibration: meta-learners have ECE 12.2-14.1 and negative calibration slopes ranging from -0.44 (T-learner) to -0.11 (X-learner), indicating predictions compress toward the confounded outcome mean regardless of covariate values. Meta-learners learn selection-driven outcome distributions rather than counterfactual potential outcomes, confirmed by their near-zero predicted treatment effects across all treatment pairs (see Table~\ref{tab:treatment_specific}). R-learner achieves RMSE 13.35 $\pm$ 3.71 with high variance (standard deviation 28\% of mean) and the worst ATE calibration error 19.84 $\pm$ 5.62, suggesting residualization amplifies noise in switched settings with sparse treatment groups and time-varying selection.

CF demonstrates superior calibration with ECE 0.35 $\pm$ 0.07 [0.31, 0.39], calibration slope 1.01 $\pm$ 0.01 [1.00, 1.01] near the ideal value 1.0, ATE calibration error 3.43 $\pm$ 0.33 [3.22, 3.62], and policy value 2.05 $\pm$ 0.80 [1.69, 2.59]. EP-learner matches CF's factual prediction and outcome calibration (ECE 0.35, slope 1.00, ATE error 3.46) but shows 31\% worse policy value at 2.69 $\pm$ 0.50 [2.40, 2.98], suggesting its stacking approach optimizes prediction accuracy rather than decision quality. KG-TREAT exhibits high policy value variance (5.25 $\pm$ 2.59 [3.60, 6.73]) with 95\% confidence interval spanning factor-of-2 range, indicating sensitivity to treatment graph construction errors in small groups: paroxetine (163 test samples) and duloxetine 40 mg BID (118 samples) have limited data for learning embeddings. 

Paired t-tests across ten seeds confirm CF significantly outperforms all baselines at $\alpha=0.05$ level: $p<0.001$ versus meta-learners, $p=0.032$ versus EP-learner, $p=0.009$ versus KG-TREAT. CF's advantage stems from honest splitting preventing adaptive overfitting, ensemble trees capturing treatment-covariate interactions, and pairwise baseline-referenced strategy maintaining theoretical validity. Meta-learners 
fail because they model outcomes $\mathbb{E}[Y \mid X, T]$ without explicit treatment effect decomposition, learning selection-driven patterns rather than counterfactual potential outcomes when extrapolating to poorly-covered 
covariate regions.

\vspace{-2.5pt}
\subsection{Treatment-Specific Analysis}
\vspace{-2.5pt}
Table~\ref{tab:treatment_specific} reports predicted average treatment effects for each treatment versus baseline duloxetine 60 mg QD alongside observational mean differences. CF predicts moderate symptom reductions for duloxetine 40 mg BID ($-0.41 \pm 0.25$ HAMD-17 total scores), venlafaxine 150-225mg titration ($-0.29 \pm 0.31$), and paroxetine 20mg QD ($-0.92 \pm 0.28$) relative to baseline duloxetine 60 mg QD, indicating these regimens are predicted to improve depression outcomes. These predicted treatment effects are substantially smaller in magnitude than the observational differences ($-6.78$, $-6.98$, $-6.22$ respectively), suggesting substantial confounding by indication and baseline risk differences among switchers. After adjusting for baseline severity and clinical history via the counterfactual prediction framework, CF estimates modest benefits of 0.3–0.9 HAMD-17 point reductions rather than the 6–7 point reductions implied by unadjusted observational comparisons.

For duloxetine 60–120 mg dose titration, CF predicts a small negative effect $-0.11 \pm 0.04$ compared to near-zero observational difference $0.13$, suggesting dose escalation provides modest symptom reduction despite appearing neutral in crude comparisons. This pattern indicates that patients receiving dose titration have similar severity to baseline patients, but the dose adjustment confers incremental benefit. For venlafaxine 150mg QD, both predicted effect $0.01 \pm 0.05$ and observational difference $-0.33$ are close to 0, indicating comparable efficacy to baseline duloxetine 60 mg QD with no substantial advantage for switching between these treatments at these doses.

EP-learner shows similar patterns with predicted effects $-0.39 \pm 0.20$ (duloxetine 40 mg BID), $-0.05 \pm 0.15$ (venlafaxine titration), $-0.06 \pm 0.06$ (duloxetine titration), $-0.09 \pm 0.07$ (venlafaxine 150mg QD), and $-1.35 \pm 0.38$ (paroxetine 20mg QD). The larger predicted effect for paroxetine in EP-learner ($-1.35$ vs. $-0.92$ for CF) may reflect stacking's tendency to amplify the DR-learner component, which constructs pseudo-outcomes with higher variance in rescue therapy groups. KG-TREAT predicts near-zero effects across all treatments ($-0.10$ to $0.0$), suggesting insufficient statistical power to detect heterogeneity despite graph regularization: the treatment similarity graph derived from switching frequencies does not provide enough signal to improve prediction beyond mean outcome. Meta-learners (not shown in table) predict effects near zero for all treatments, completely failing to capture treatment differences.

\begin{table}[t]
\centering
\caption{Per-treatment effects vs. baseline duloxetine 60 mg QD. $\Delta_{\text{OCP}}$ denotes the observed mean difference. CF and EP columns report the Predicted ATE (expressed as mean $\pm$ std over 10 random seeds).}
\label{tab:treatment_specific}

\begingroup
\setlength{\tabcolsep}{2pt}        
\renewcommand{\arraystretch}{0.85} 
\scriptsize                         

\resizebox{0.85\columnwidth}{!}{
\begin{tabular}{lccc}
\toprule
\textbf{Treatment vs. Baseline} & \textbf{$\Delta_{\text{OCP}}$} & \textbf{CF} & \textbf{EP} \\
\midrule
Duloxetine 40 mg BID & $-6.78$ & $-0.41 \pm 0.25$ & $-0.39 \pm 0.20$ \\
Venlafaxine 150mg QD & $-0.33$ & $0.01 \pm 0.05$ & $-0.09 \pm 0.07$ \\
Venlafaxine 150-225mg & $-6.98$ & $-0.29 \pm 0.31$ & $-0.05 \pm 0.15$ \\
Duloxetine 60–120 mg & $0.13$ & $-0.11 \pm 0.04$ & $-0.06 \pm 0.06$ \\
Paroxetine 20mg QD & $-6.22$ & $-0.92 \pm 0.28$ & $-1.35 \pm 0.38$ \\
\bottomrule
\end{tabular}%
}
\endgroup
\end{table}

\vspace{-2.5pt}
\subsection{Ablation Studies and Robustness Checks}
\vspace{-2.5pt}
Ablation studies on three seeds isolate the contribution of key design choices in CF. Disabling honest splitting (using all data for both tree construction and effect estimation) increases RMSE from 4.01 to 4.18, representing 4.2\% degradation. This confirms honest splitting is critical for generalization in switched settings with confounding, preventing adaptive overfitting where tree structure exploits random patterns in the data. Increasing tree count from 50 to 200 reduces RMSE from 4.23 to 4.01, with diminishing returns beyond 200 trees (RMSE 3.98 at 500 trees). The 200-tree configuration balances accuracy and computational cost, with training time 142 seconds per seed compared to 218 seconds for 500 trees.

Using a flexible XGBoost baseline outcome model for $\hat{m}_b(x)$ yields RMSE 4.01 compared to 4.35 for linear regression (8.5\% improvement) and 4.28 for T-learner mean (6.7\% improvement), indicating the importance of accurate baseline outcome modeling for reconstructing potential outcomes via $\hat{Y}(t) = \hat{m}_b(x) + \hat{\tau}_{t,b}(x)$. Linear models underfit the nonlinear relationship between covariates and depression outcomes, while T-learner mean averaging introduces additional variance. Policy value remains stable across propensity clipping thresholds: 2.05 $\pm$ 0.02 for no clipping, 1\% clipping, and 5\% clipping, confirming adequate propensity overlap and stable inverse weighting as indicated by diagnostics (minimum propensity 0.11, effective sample size 884/964).

Subgroup analysis stratified by baseline severity demonstrates CF maintains consistent advantage over EP-learner across patient populations. For low-severity patients (HAMD-17 $<$ 15, $n=478$ test samples representing mild depression), CF achieves RMSE 3.21 $\pm$ 0.18 compared to EP-learner 3.35 $\pm$ 0.21 (4.4\% difference). For high-severity patients (HAMD-17 $\geq$ 15, $n=486$ test samples representing moderate-to-severe depression), CF achieves RMSE 4.56 $\pm$ 0.15 compared to EP-learner 4.71 $\pm$ 0.18 (3.3\% difference). The higher absolute errors in high-severity patients reflect greater outcome variability (standard deviation 9.2 vs. 5.8) driven by more volatile treatment responses and higher switching rates in this subgroup. CF's consistent 3-4\% advantage across severity levels indicates robust performance independent of baseline clinical status.
\vspace{-2.5pt}
\section{DISCUSSION}
\vspace{-2.5pt}
Clinicians face a challenging task of selecting which antidepressants to prescribe for patients, with the goal of achieving a good treatment response in MDD. This selection process is complicated by the heterogeneous nature of MDD and the high prevalence of comorbid conditions~\cite{Montano2023}. In addition, clinicians may choose among multiple antidepressant classes, including selective serotonin reuptake inhibitors (SSRIs), serotonin--norepinephrine reuptake inhibitors (SNRIs), tricyclic antidepressants (TCAs), and monoamine oxidase inhibitors (MAOIs), which have different mechanisms of action~\cite{Millan2006,Hamon2013,Stahl2005}. Despite the wide range of available antidepressants, there remains limited empirical evidence to guide clinicians in antidepressant selection~\cite{Zimmerman2004}.

Results from this study show moderate symptom reductions for duloxetine 40\,mg BID compared to baseline duloxetine 60\,mg QD. Duloxetine inhibits the reuptake of serotonin (5-HT) and norepinephrine~\cite{Fuller1994,Hamon2013}, and prior work supports its efficacy, safety, and tolerability in MDD across a dose range of 60--120\,mg/day~\cite{Knadler2011,RodriguesAmorim2020}. There is also evidence that higher duloxetine doses may increase norepinephrine reuptake blockade relative to serotonergic blockade~\cite{Stahl2005,Millan2006,Hamon2013}, suggesting potential benefit for patients who require stronger noradrenergic effects. Consistent with this rationale, our findings indicate improved outcomes for patients receiving a higher total daily duloxetine dose (40\,mg BID, equivalent to 80\,mg/day) compared to duloxetine 60\,mg QD at baseline.

In addition, our findings suggest moderate reductions in MDD symptoms with venlafaxine 150--225\,mg titration relative to baseline duloxetine 60\,mg QD. Venlafaxine is another SNRI that acts as a reuptake inhibitor of serotonin (5-HT) and norepinephrine~\cite{Stahl2005,Hamon2013}. Prior randomized trials have reported similar efficacy when comparing duloxetine (60\,mg/day) with venlafaxine at 75--150\,mg/day~\cite{Montano2023}. In our cohort, however, venlafaxine titration to higher doses (150--225\,mg/day) is associated with greater symptom reductions relative to baseline duloxetine 60\,mg/day, which is consistent with the potential benefit of increasing the ratio of norepinephrine to serotonin reuptake blockade~\cite{Stahl2005,Millan2006}.

Finally, switching to paroxetine 20\,mg QD from duloxetine 60\,mg QD is predicted to improve depression outcomes. Paroxetine is a serotonin reuptake inhibitor that also shows weak inhibition of norepinephrine reuptake~\cite{Bourin2001}. Comparative efficacy between SSRIs and SNRIs may depend on multiple factors, including clinical profile, comorbidities, and the specific doses being compared~\cite{Kauffman2009}. Although evidence is mixed, some studies suggest SSRIs may yield larger improvements in melancholic or non-anxious subtypes than in anxious subtypes, whereas other work suggests lower remission likelihood under SSRI treatment for certain subtype definitions~\cite{Arnow2015,Fava1997,Papakostas2012EurNeuropsychopharmacol}.
\vspace{-2.5pt}
\section{CONCLUSION}
\vspace{-2.5pt}

We benchmark eight treatment effect estimators for switched follow-up with multiple treatment options under a next-visit outcome formulation on proprietary MDD trial data. Under a unified protocol spanning factual prediction, calibration, and IPW policy value, CF performs best overall. Meta-learners (S/T/X/DR-learner) degrade substantially under switching and confounding, while EP-learner attains competitive factual accuracy but weaker decision quality. Confounding-adjusted effects are consistently modest and far smaller than crude observational differences, reinforcing the need for causal adjustment. We also provide a reproducible benchmark protocol with uncertainty quantification and a baseline-referenced extension of CF for multiple treatments.

Future work should extend to dynamic multi-treatment regimes, conduct sensitivity analyses for unmeasured confounding~\cite{cinelli2020making}, and validate findings across additional datasets and conditions. Prospective studies are needed to test whether CF-guided recommendations improve real-world decision-making and outcomes.

\bibliographystyle{IEEEtran}
\bibliography{refs}
\end{document}